# Skipper: Asynchronous Maximal Matching with a Single Pass over Edges

Mohsen Koohi Esfahani
0000-0002-7465-8003

***Abstract*—Maximal Matching (MM) is a fundamental graph problem with diverse applications. While state-of-the-art parallel MM algorithms have a total expected work linear in number of edges, they require randomization, iterative graph processing, and graph pruning after each iteration. These overheads increase execution time and demand additional memory, reducing applicability to large-scale graphs.**

**In this paper, we introduce *Skipper*, an asynchronous Maximal Matching algorithm that resolves conflicts *instantaneously* using a parallel reservation strategy, which merges both reservation and committing steps into a single step.**

**Skipper processes each edge only once, definitively determining whether the edge is selected as a match. Skipper does not require graph pruning and minimizes memory space utilization, requiring only a single byte of memory space per vertex. Furthermore, Skipper operates in the asynchronous parallel random access machine (APRAM) model, relaxing synchronization between threads, and facilitating better parallelization gains.**

**Our evaluation, conducted on real-world and synthetic graphs with up to 224 billion edges, shows that Skipper achieves a speedup of 4.9–15.6 times, with a geometric mean of 8.0 times.**

***Index Terms*—Maximal Matching, Shared-Memory Graph Algorithms, Parallel Reservation**

## I. Introduction

Maximal Matching (MM) is a core graph problem with a wide range of applications in science, industry, and humanities, including optimizing resource allocation, pairing in biological networks, and analysis of pairwise collaboration in social networks. The goal of MM is to find a maximal set of edges that do not share endpoints. In other words, a subset of the graph edges is maximal if every edge in the graph shares at least one endpoint with an edge in the set.

A sequential MM algorithm can be implemented using a greedy approach. This algorithm, referred to as **Sequential Greedy MM** (SGMM), maintains a *status* array to track whether a vertex has been matched to an edge. The algorithm iterates over all edges of the graph, and for each edge, if both endpoints are unmatched, the edge is selected and the *status* of its endpoints is updated to matched.

In a parallel setting, multiple threads process edges concurrently. If two or more threads concurrently select edges with common endpoints, it can lead to **conflicts**, resulting in an incorrect outcome. This necessitates finding solutions to manage conflicts in parallel MM.

The primary approach employed in state-of-the-art MM algorithms [1–6] is **Endpoints' Mutual Selection** (EMS). In EMS, vertices independently select a neighbor, and if two vertices have mutually selected each other, their connection is selected as a match. Since a selected neighbor may select its other neighbor, some neighboring vertices may be left with no matched edges. Consequently, to ensure maximality, EMS requires **iteratively** processing vertices and removing the matched vertices and their incident edges from consideration in subsequent iterations, i.e., **graph pruning**[1].

It has been shown that by **randomizing** the selection of vertices, it is guaranteed that the number of unmatched vertices decreases geometrically in each iteration, resulting in a total expected work that is linear in number of edges [1–3].

The main performance limitations of EMS-based algorithms in processing real-world graphs are: (i) iterative processing of unmatched vertices/edges, (ii) graph pruning in each iteration, which increases memory demand and execution time, and (iii) the overhead of randomization.

To mitigate the overhead of processing all remaining edges in each iteration, Prefix-Batched MM (PBMM) proposes to process a subset of edges in each iteration [3]. Similarly, to alleviate the overhead of randomization, Sampling-based Internally-Deterministic MM (SIDMM) [7], iteratively samples a subset of the remaining edges in the graph.

Despite the success of previous studies in parallelizing MM computation, their execution is still dependent on iterative graph processing, graph pruning, and randomization. This has resulted in inefficient utilization of computing resources.

Our evaluation reveals that parallel execution of SIDMM on 64 threads, requires 33–58 times as many memory accesses as the sequential SGMM requires. This significant overhead leads to **limited parallelization gains**. Specifically, SIDMM achieves parallelization gains ranging from 1.7 to 4.5 times compared to SGMM.

In this paper, we introduce **Skipper**, a parallel asynchronous MM algorithm. In contrast to EMS that accumulates all conflicts and tries to resolve them in the second step, Skipper introduces **Just-In-Time (JIT) conflict resolution** which resolves the conflicts as soon as they appear. The JIT conflict resolution processes both endpoints of each edge simultaneously, allowing it to determine whether the edge is a match or not in **a single, coordinated step**. By the end of processing an edge, it is either (1) selected as a match or (2) at least one of its endpoints is matched to another vertex by another thread. Consequently, there is *no need to reconsider* this edge in the future.

[1]Note that "pruning" refers generally to the bookkeeping that removes selected edges and their adjacent edges, either through materialized deletion or through other, probably more efficient methods.

Skipper's JIT conflict resolution allows processing the graph in a **single pass over the edges**, without requiring pruning or randomization. Moreover, we show that JIT conflicts are rare and the conflict resolution has minimal impacts on performance and work-efficiency.

In Skipper, a thread continuously processes edges by only accessing the data structures shared between threads and the topology data of the edge being processed. This facilitates Asynchronous Parallel Random Access Machine (**APRAM**) model for Skipper.

The total expected work complexity of Skipper is linear in the number of edges. In terms of memory space utilization, Skipper requires only one byte of memory space per vertex. While efficient, Skipper is simple, with no tuning parameters, and can be integrated into a graph processing framework with fewer than 50 lines of code.

We evaluate Skipper in comparison to SIDMM, a state-of-the-art MM, using real-world and synthetic graphs with up to 224 billion edges, where Skipper achieves a significant speedup, ranging from 4.9 to 15.6 times, with a geometric mean of 8.0 times. Notably, for graphs larger than 50 billion edges, the geometric mean speedup increases to 10.3 times.

The main contributions of this paper are:

- Analyzing previous MM algorithms and evaluating their work efficiency and parallelization gain,
- Identifying the main limitations of the EMS method in resolving conflicts,
- Introducing Skipper, an asynchronous MM algorithm that requires a single pass over edges,
- Presenting a theoretic analysis of Skipper, and
- Experimentally evaluating Skipper in comparison to a state-of-the-art MM algorithm.

The paper is continued with a review of background materials and related work in Section II. Section III investigates the sources of inefficiency in the state-of-the-art MM algorithms. We introduce Skipper in Section IV which is analyzed in Section V and evaluated in Section VI. Section VII concludes the discussion.

## II. Background & Related Work

### A. Terminology

The immutable undirected graph $G = (V, E)$ consists of a set of vertices $V$, and a set of edges $E$ between these vertices. Edges are unordered pairs of elements of $V$. The set of neighbors of vertex $v$ is denoted by $N_v$.

The Compressed Sparse Row (CSR) or Column (CSC) [8, 9] format is a compact representation for sparse matrices and graphs. It represents a graph using two arrays:

- the **offsets** array containing $|V| + 1$ elements, and
- the **neighbors** array of $|E|$ elements.

The `offsets` array is indexed by a vertex ID and specifies the index of the first edge of that vertex in the `neighbors` array. The `neighbors` array specifies the ID of the *source* endpoints of the edges in CSC and the *destination* endpoints in CSR. In a symmetric graph the CSC and CSR presentations are equivalent. $N_v$ is defined as:

$$N_v \;=\; \{\; \text{neighbors}[i] \mid \text{offsets}[v] \;\leq\; i \;<\; \text{offsets}[v+1] \;\}$$

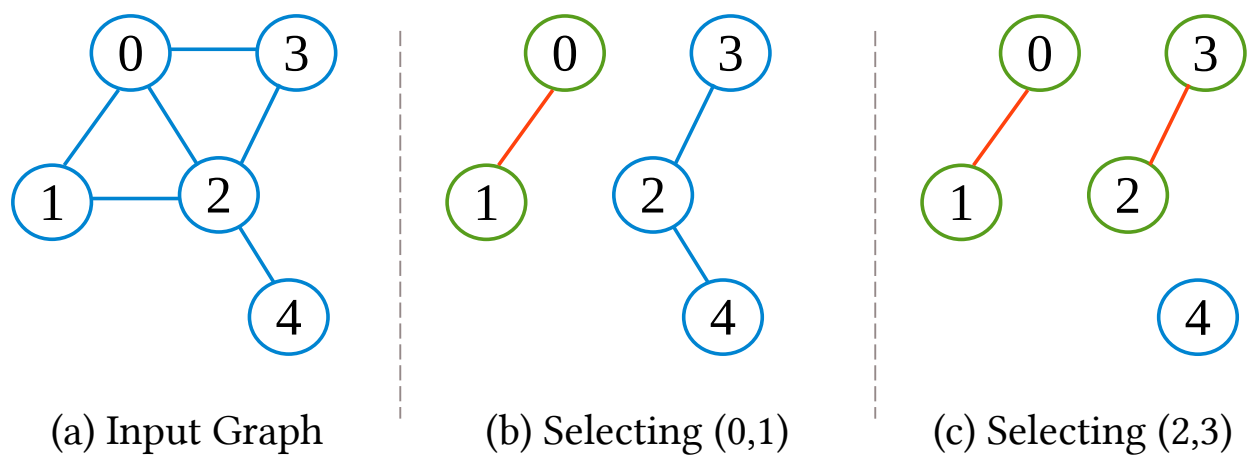

Fig. 1: An example graph (a), and two steps of execution of SGMM (b,c). In (b), edge (0,1) is selected (orange line) and its endpoints, i.e. vertices 0 and 1, are marked (green borders). In (c), edge (2,3) is selected and its endpoints are marked.

### B. Maximal Matching

In **Maximal Matching (MM)**, the goal is to identify a maximal subset of $E$ containing no common endpoints. The MM output is validated by checking that (a) each graph edge has at least one common endpoint with an edge in the output and (b) no two edges in the output share an endpoint.

The **Sequential Greedy MM (SGMM)** iterates over edges. For each edge, if neither endpoints is already marked, the edge is added to the output, and the endpoints are marked to prevent being selected later. In this way, SGMM requires a single bit of memory space per vertex.

Alternatively, a sequential greedy approach can be used where all edges incident to the endpoints of a selected edge are removed from the graph. This approach eliminates the need for memory allocation per vertex but incurs a higher cost for graph pruning. Consequently, the former version has a better performance, and we use it when referring to SGMM.

When the graph is processed in CSR format by SGMM, after selecting an edge as a match, the next neighbors of the current vertex do not need to be processed. As shown in Section VI-C, the total number of memory accesses in SGMM is 0.3–0.8 times the number of graph edges.

Figure 1 illustrates an example graph and the execution of SGMM. Starting from vertex 0, edge (0,1) is the first selected edge as a match (orange line in Figure 1.b) and its endpoints are marked (green borders in Figure 1.b). As a result, other edges incident to vertex 0 do not need to be processed. The next vertex is 1, which is already marked and the execution is continued by processing vertex 2. Since edges (2,0) and (2,1) cannot be selected as their other endpoint being marked, edge (2,3) is selected as a match (Figure 1.c).

### C. Endpoints Mutual Selection

When processing a graph using parallel threads, two or more threads may concurrently process edges with dependencies on each other, potentially leading to **conflicts**.

For example, consider the graph in Figure 1.a. If thread $t_0$ processes edge (0,1) and thread $t_1$ concurrently processes edge (3,0), they may both observe that vertex 0 is unmarked. Consequently, they select these edges as matches, despite sharing a common endpoint, which would invalidate the correctness of the output.

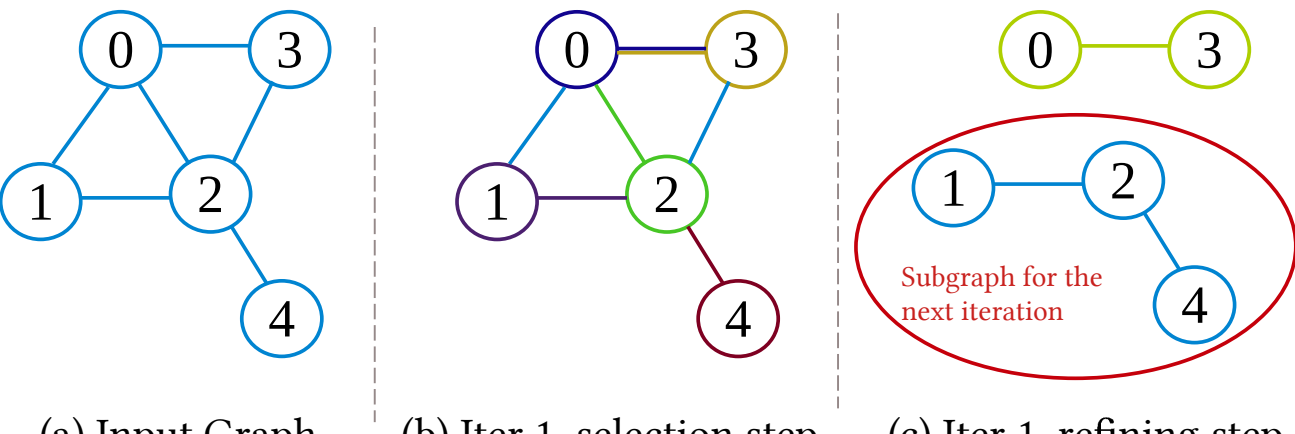

(a) Input Graph (b) Iter 1, selection step (c) Iter 1, refining step

Fig. 2: A sample execution of one iteration of the EMS method. The example graph (a) is processed in two steps. In the selection step (b), each vertex selects one of its edges. We used the same color for the selected edge by a vertex. Vertices 0 and 3 selected edge (0,3), vertex 1 selects its edge to 2, and so on. In the refinement step (c), mutually selected edges by their endpoints are chosen as matches. Only edge (0,3) is selected by both endpoints and is chosen as a match. The graph after removing matched vertices and their edges (subgraph in bottom of (c)) is processed in the next iteration.

To manage conflicts in parallel MM, State-of-the-art algorithms primarily rely on the **Endpoints Mutual Selection (EMS)** method. The core idea is to leverage the MM requirement that each vertex appears in the output at most once. The EMS requires two rounds:

- The *“selection step”*, where each vertex selects one of its edges as its candidate match.
- The *“refinement step“*, where the candidate edges that mutually have been selected by their both endpoints are chosen as matches.

Figure 2 illustrates an example execution of the EMS method. Initially, each vertex selects one of its edges (Figure 2.b). Then, mutually selected edges by both endpoints are chosen as matches (Figure 2.c). As shown in Figure 2.c, only edge (0,3) is selected in this iteration.

To ensure **maximality**, EMS requires **multiple iterations**. After each iteration, **graph pruning** is performed by removing the matched vertices and their incident edges. The next iteration is executed on the updated graph. By **randomizing** the selection of vertices or edges, the number of unmatched vertices decreases geometrically and the total expected work becomes linear in the number of edges [1–3].

### D. Parallel MM Algorithms

The previous studies on parallel MM are mainly based one the EMS method Section II-C. In this section, we review their differences in applying EMS.

Israeli and Itai [1] introduce an iterative MM algorithm. In each iteration, each vertex selects a random edge. If both endpoints have selected the same edge, the edge is added to the output. The selected endpoints and their neighbors are removed from the graph before passing to the next iteration.

Auer and Bisseling [2] propose an iterative EMS algorithm where each iteration involves randomly assigning red or blue labels to unmatched vertices. Blue vertices then propose matching to their red neighbors, and the red neighbors select one proposal which is added to the matches. The vertices that are blue or red and receive/send no proposal are removed from consideration in the next iteration.

Blelloch et al. [3] present the *Prefix-Batched MM (PBMM)* algorithm. It takes a fixed random priority on vertices as input. In each iteration, edges with no adjacent edges of higher priority are selected as matches. Since the order of edges is resolved by the input priority, the algorithm output is deterministic.

To reduce redundant work, PBMM employs **prefix batching** to bound the number of elements processed per iteration. Each iteration processes a set of unmatched edges carried over from the previous iteration, plus a new batch of previously unprocessed edges. Prefix batching uses a “granularity” parameter to trade off between parallelization and work efficiency.

Blelloch et al. [4] present the *Internally-Deterministic MM (IDMM)* which performs the EMS in a **parallel reservation** paradigm. IDMM assigns a unique ID to each edge and operates in two phases per iteration. In the *“reserve”* phase, each vertex updates its data with the minimum ID of its incident edges. In the *“commit”* phase, edges with matching IDs indicated by both endpoints are selected as matches. Edges with at least one matched endpoint are excluded from consideration in subsequent iterations. The resolution of matches based on the ID of edges results in a deterministic output. Additionally, prefix-batching is applied to IDMM to enhance its performance.

Birn et al. [5] propose an EMS algorithm that assigns random weights to edges in each iteration. Then, each vertex selects the heaviest edge incident on it. If two vertices have selected the same edge, this edge is added to the matches. All edges of the matched endpoints are removed from the graph, and the pruned graph is passed to the next iteration.

Lim and Chung [6] propose a **distributed EMS** algorithm which begins an iteration with broadcasting the degree of unmatched vertices to their unmatched neighbors. Then, each neighbor selects the neighbor with the lowest degree and sends it a match message. If a vertex receives a match message from the neighbor it had sent a message to, their link is selected as a match. The degree of a vertex may decrease in an iteration as some vertices become matched and inactive.

To reduce the overhead of randomization, **Sampling-based IDMM (SIDMM)**[2] in the Graph-Based Benchmark Suite (GBBS) [7], iteratively samples a subset of remaining edges of the graph and processes them using IDMM. The main idea is that we only need randomized selections among unmatched vertices. Therefore, it is not necessary to pay the cost of randomizing the whole graph.

SIDMM performs the sampling in two passes over vertices in each iteration. In the first pass, it builds a offsets array from the number of unmatched neighbors per vertex which is used to pick sample positions. The second pass maps the sampled positions to source vertices and scans only the necessary neighbor lists to retrieve the live edges.

[2]The algorithm is referred to by “Random Greedy MM” in the framework, which can also be conceptually applied to a set of other MM algorithms. To better distinguish it, we refer to it as “SIDMM”.

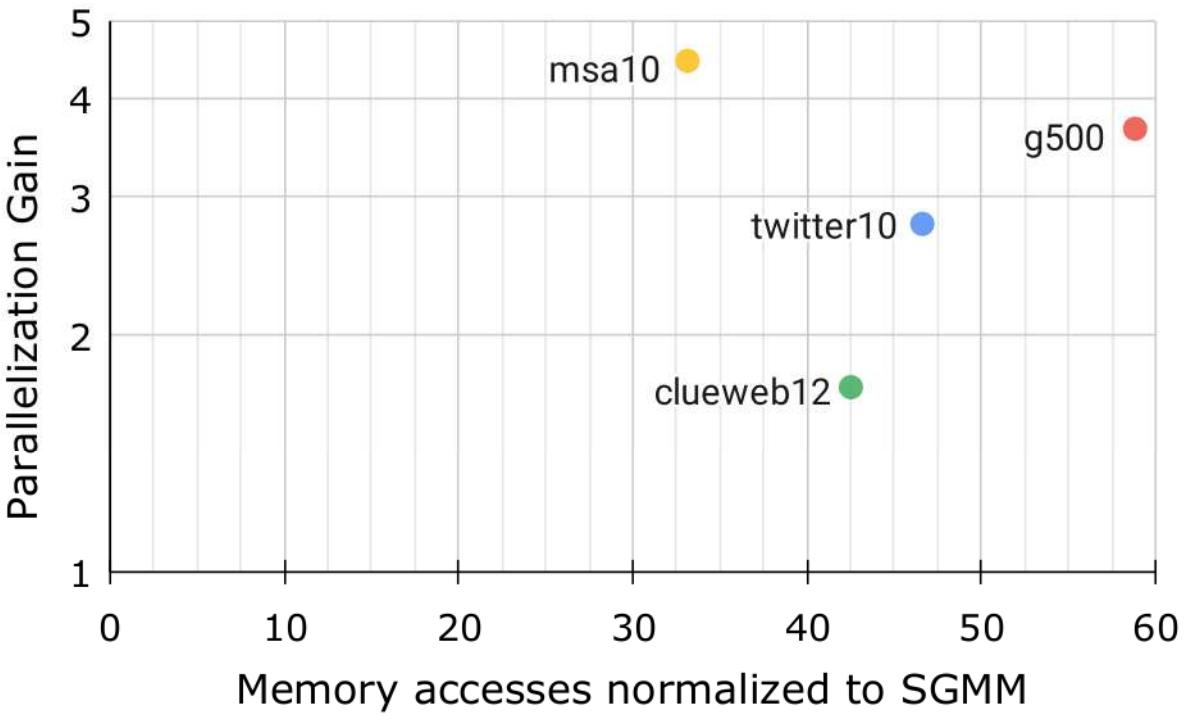

Fig. 3: Comparing performance and memory overhead of SIDMM, a parallel MM running on 64 threads, with SGMM running on a single thread. The y-axis represents the Parallelization Gain, i.e., the execution time of SGMM divided by SIDMM. The x-axis shows the total number of memory accesses (i.e., load and store instructions) of SIDMM, normalized to the number of memory accesses performed by SGMM. Datasets and the machine are explained in Section VI.

Consequently, while the subgraph is not materialized, the effects of graph pruning and randomization are achieved via sampling in SIDMM. Sampling is controlled by a parameter that specifies the number of samples per iteration.

## III. Motivation

In Section II-D, we explained that state-of-the-art MM algorithms facilitate parallelization, they require iteratively processing the graph, randomization, and graph pruning. These requirements impose significant overhead on the performance and reduce work efficiency.

To better consider the impacts of these overheads on the execution of parallel MM algorithms, we measure the work efficiency and performance of SIDMM (a parallel MM) with SGMM (a sequential MM). Since MM for large-scale graphs is a memory-bound application, we use the number of memory accesses to measure the work efficiency of previous works.

Figure 3 compares SIDMM utilizing 64 threads[3], to SGMM as a sequential MM. The y-axis shows the parallelization gain, calculated is defined as the execution time of SGMM divided by that of SIDMM. The x-axis shows the total number of memory accesses in SIDMM normalized by the memory access in SGMM. Datasets and the machine are detailed in Section VI.

Figure 3 shows that SIDMM performs 33–58 times as many memory accesses as SGMM, with a geometric mean of 44 times. This significant overhead leads to **limited parallelization gains**. The figure shows that, SIDMM achieves a parallelization gain of 1.7–4.5 times compared to SGMM, with a geometric mean of 3.0 times.

In other words, parallelization in SIDMM increases the total work by a factor of 44 times, but only improves performance by a factor of 3. This highlights that more efficient parallel MM algorithms are required to better utilize computing resources.

To identify the primary sources of the above limitations, we take a closer look at the EMS method, which reveals the main reason of inefficiency: EMS processes the endpoints of an edge **independently** from each other, and in two **separate** passes over the graph: a first pass to select the candidates and a second pass to resolve the conflicts and determine acceptable candidates as matches. Since the initial selection may be canceled in the second pass, the iterative execution becomes necessary and the overheads of randomization and graph pruning are imposed.

This suggests that by **combining the computation of both endpoints of an edge into a single definitive step**, we can eliminate the limitations of EMS and improve the work efficiency and parallelization gain. This is the main idea behind designing our parallel Maximal Matching algorithm, Skipper, which is introduced in the next section.

## IV. Skipper Maximal Matching

### A. Algorithm Design

In Section III, we explained that the EMS's requirement to separately process endpoints of edges has made significant overhead and resulted in in poor parallelization gains.

To avoid the EMS's overheads, we need to process both endpoints of an edge in a single step, eliminating a second pass to solve conflicts. In other words, while concurrent threads process different edges that may conflict, we need a method for threads to **resolve a conflict on the moment it is made and before further progressing, i.e., Just-In-Time (JIT) conflict resolution**. If successful in designing a JIT conflict resolver, after processing an edge, it is **irreversibly decided** whether that edge is a match or not (since one of its endpoints have been matched by another thread). Consequently, we can process the graph in a *single pass over edges* without requiring iterative graph processing, graph pruning, and randomization.

To achieve this goal, we introduce our parallel MM algorithm, **Skipper**, which adopts a **parallel reservation strategy** to resolve JIT conflicts efficiently. Skipper assigns a *state* value to each vertex. A vertex can be in one of three states: `accessible`, `reserved`, and `matched`. Matching is initiated by changing the state of the first endpoint from `accessible` to `reserved` and attempting to change the state of the second endpoint to `matched`. If successful, the state of the first endpoint is also changed to `matched`.

Using the `reserved` state, we manage the **temporary uncertainty** in the state of the first endpoint and instruct other threads to wait until the state becomes certain. In this way, while at least one thread progresses, dependent threads wait and, after some attempts, will have the certain state (`accessible` or `matched`) and continue their work.

[3]Note that, as we explain in Section VI-D, utilizing $t$ threads ($t > 1$) in memory-intensive applications does not directly translate to accessing $t$ times the computational resources available to a single thread.

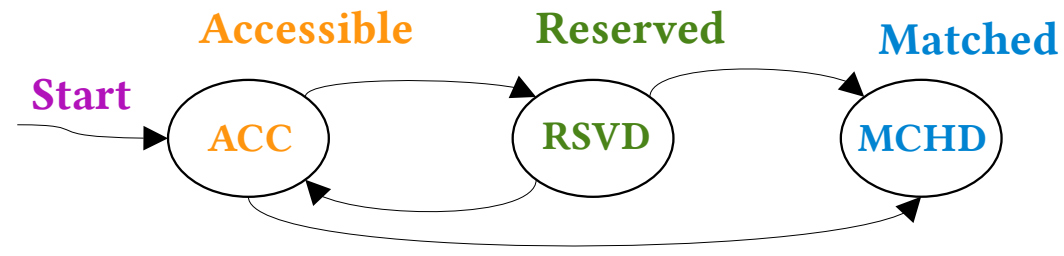

Fig. 4: State transition diagram of a vertex in Skipper

By efficiently resolving JIT conflicts, Skipper avoids EMS's strategy of accumulating all conflicts and resolving them in a separate step. Consequently, the whole graph is processed in a single pass over edges, eliminating the EMS's overheads. Since JIT conflicts are rare (analyzed in Section V-B and experimentally evaluated in Section VI-E), Skipper achieves a high performance and work efficiency.

### B. High-Level Overview

To track the matching status of a vertex and manage conflicts, Skipper assigns a *state* value to each vertex. A vertex can be in one of three states:

- **Accessible (ACC)**: The vertex is unmatched and accessible to all threads.
- **Reserved (RSVD)**: The vertex is temporarily reserved for a thread and inaccessible to other threads. The vertex will eventually transition to either the `matched` or `accessible` state. The state can be changed only by the thread holding the reservation.
- **Matched (MCHD)**: One edge of the vertex has been selected as a match and it is permanently inaccessible.

Throughout our discussion, we use *"releasing"*, *"reserving"*, or *"matching"* a vertex to refer to changing its state to `ACC`, `RSVD`, and `MCHD`, respectively. Figure 4 illustrates the state transition diagram of a vertex in Skipper. Initially, all vertices are in `ACC` state, from which they can transition to either `RSVD` or`MCHD` states. In the `RSVD` state, a vertex can return to `ACC` or transition to `MCHD`. The `MCHD` state is an irreversible, meaning that once a vertex enters this state, it cannot transition to other states.

The matching process for an edge in Skipper begins by checking if neither endpoint is matched (i.e., not in the `MCHD` state) and attempting to reserve the first endpoint (i.e., changing the state to `RSVD`). If multiple threads try to reserve a vertex, only one thread will successfully change the state and proceed, while others wait for the result.

After reserving the first endpoint, the thread attempts to match the second endpoint. If successful, the thread sets the state of the first endpoint to `MCHD`. The waiting threads observe this change and recognize that the vertex is no longer accessible, moving on to process other edges. On the other hand, if the thread fails to match the second endpoint, the first endpoint is released (i.e., returning to `ACC`), allowing the waiting threads to reserve or match it.

The `RSVD` state provides flexibility in two ways:

- It temporarily prevents concurrent threads from changing the state of the first endpoints, allowing a single thread to proceed and finalize its second endpoint's state,

**Algorithm 1:** Skipper Maximal Matching

```
Input: G(V, E)
   /* Initialization                                              */
 1 matches[ ];
 2 state[|V|];
   /* State values of a vertex:
        0: ACC, 1: RSVD, 2: MCHD                                   */
 3 par_for v ∈ V
 4 |  state_v = ACC;

   /* Traversing edges                                            */
   /* Thread-dispersed locality-preserving scheduling             */
 5 par_for e(x, y) ∈ E
      /* Skip loops                                               */
 6    if x == y then
 7    |  continue;
      /* Prevent cycles                                           */
 8    u = min(x, y);
 9    v = max(x, y);

      /* As long as no endpoint is matched                        */
10    while state_u ≠ MCHD ∧ state_v ≠ MCHD do
         /* Try setting u as reserved                             */
11       if ! CAS(state_u, ACC, RSVD) then
12       |  continue;
         /* Try setting v as matched                              */
13       while state_v ≠ MCHD do
14          if CAS(state_v, ACC, MCHD) then
               /* Set u as matched                                */
15             state_u = MCHD;
               /* Add (u, v) to matches                           */
16             matches.push((u, v)) ;                    // Race-free
17       if state_u ≠ MCHD then
            /* Unsuccessful match. Release u.                     */
18          state_u = ACC;

19 return matches;
```

- It allows waiting threads to continue after the reserving thread completes processing its edge.

Using this mechanism, Skipper temporarily reduces parallelism while concurrent threads process conflicting edges to achieve just-in-time (JIT) conflict resolution. The overhead for each conflict is just a few CPU cycles spent in waiting for the reserve holder to complete its processing, without imposing iterative graph processing. Since JIT conflicts occur rarely (Section V-B and Section VI-E), **Skipper maintains normal execution with high parallelism and work efficiency and only reduces parallelism in the worst-case scenarios (when JIT conflicts happen)**.

### C. Pseudo Code

Algorithm 1 shows the pseudo code of Skipper. Skipper uses a *state* array of size $|V|$, where each vertex can be in one of three states: $ACC(0)$, $RSVD(1)$, and $MCHD(2)$ denoting `accessible`, `reserved`, and `matched`, respectively. The initial state of each vertex is set to $ACC$ (Lines 3–4).

To facilitate concurrent updates to the $state$ variable, Skipper utilizes the *Compare-And-Swap (CAS)* instruction, which atomically compares and updates the contents of a memory address. *CAS*$(X, x_{old}, x_{new})$, atomically checks the contents of $X$ and if it is equal to $x_{old}$, updates it with $x_{new}$, returning $true$. Otherwise, *CAS* returns $false$. Among multiple threads concurrently executing *CAS* on the same variable ($X$) and for the same old value ($x_{old}$), only a single thread receives $true$ as the output of *CAS*.

Edges are processed concurrently by parallel threads (Lines 5–18). To minimize the conflicts, a **thread-dispersed locality-preserving** scheduling method is employed. In this approach, the graph is divided into a set of blocks of consecutive vertex/edge IDs, with each block having approximately the same number of edges. The blocks are then assigned to threads in a contiguous manner, ensuring that threads process consecutive blocks of vertices, while being dispersed across the graph. When a thread completes its assigned blocks, it engages in work-stealing.

The *thread-dispersed* and *locality-preserving* features of this approach reduce the JIT conflicts by (i) having threads process independent neighborhoods and (ii) processing vertices with dependencies (i.e., in the same neighborhood) sequentially by a thread. For low locality and randomized graphs, randomization guarantees minimized JIT conflicts.

Lines 6–7 skip processing self-edges. Lines 8–9 find the endpoints with lower and greater IDs to prevent deadlocks in reserving vertices. As an example, a reservation deadlock may occur when thread $t_0$ is processing edge (a,b) while $t_1$ is processing edge (b,a). If $t_0$ and $t_1$ simultaneously reserve a and b, respectively, then their attempts to match the other endpoint results in an infinite wait, i.e., a deadlock. Similar situations may occur when three threads simultaneously process vertices (a,b), (b,c), and (c,a).

Lines 10–18 contain the loop for processing edge $(u,\ v)$, where $u < v$. This edge is considered only if neither of its endpoints is in `MCHD` state (Line 10).

The first step is to reserve $u$ (Lines 11–12) by atomically changing its state to `RSVD`, making it temporarily inaccessible to other threads. The next step is to change the state of $v$ to `MCHD` (Lines 13–16). This involves a $while$ loop that repeatedly attempts to modify $state_v$ (Line 14). If successful, $state_u$ is also set to `MCHD` (Line 15), and the edge $e(u,\ v)$ is selected as a match and added to the $matches$ array (Line 16).

If another thread successfully sets $state_v$ to `MCHD`, the current thread releases $u$ by changing its state to `ACC` (Lines 17–18), making it accessible to other threads. Note that *CAS* is not required in Lines 15 and 18, as $u$ has been exclusively reserved for the thread holding the reservation, and no other thread can change $state_u$ while it is reserved.

To manage **concurrent accesses of threads to the** $matches$ **array** and writing the found matches (Line 16), per-thread buffers are assigned. To reduce system calls for memory allocation, a memory block of $|V|$ edges is initially allocated. Each thread requests a 1024-edge buffer and writes its results in this buffer. This approach ensures that each thread has a private buffer which is sequentially written, without any access from other threads. When a buffer is filled, a new one is requested. The last buffer of each thread may have unfilled elements, which are filled with $-1$ to indicate invalid values. The filled buffers are used to indicate the $matches$ array and are returned as the output of the function. It can be easily processed in parallel/sequentially by skipping from invalid elements.

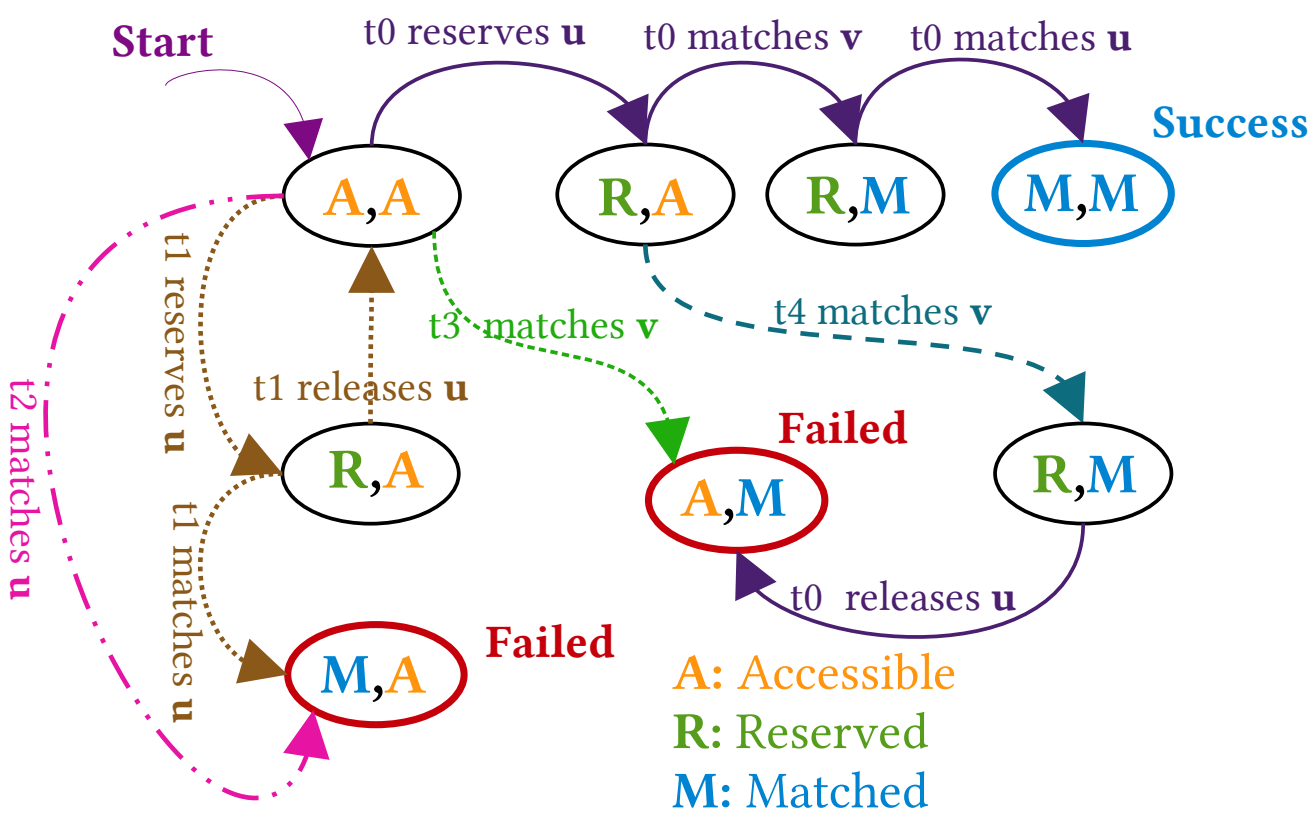

Fig. 5: An example of state transitions of endpoints in Skipper. The edge $(u, v)$, where $u < v$, is processed by thread t0 while threads t1–t4 are concurrently processing other edges of $u$ and $v$. Each node in the diagram shows two states, corresponding to the states of u and v. To improve clarity, we replicated states (R,A) and (R,M) based on the thread holding the reservation and the thread has successfully matched v. Note that the diagram omits some states and transitions for brevity. For example, it does not depict the transition from (A,A) to (A,R), where a thread reserves v.

### D. Examples

**State Transitions of an Edge.** Figure 5 illustrates an example of how states of two vertices $u$ and $v$, where $u < v$, as endpoints of the edge $(u, v)$ may change when this edge is processed by thread t0, while other edges of $u$ and v are concurrently processed by threads t1–t4. Each state is denoted by its first letter. Figure 5 shows three final states: (M,M), (A,M), and (M,A). The state (M,M) indicates that t0 has successfully processed edge $(u, v)$ and added it to the matches. States (A,M) and (M,A) indicate another thread has matched one of the endpoints, making it impossible to select edge $(u, v)$ as a match.

For thread t0 (dark purple lines) to successfully match $u$ and v, it must start by transitioning from state (A,A) to (R,A), indicating the reservation of u. Then, t0 must transition to (R,M) showing $v$ as matched, and finally to (M,M) to match u. However, when $u$ and $v$ are in state (A,A), thread t1 (brown lines) may successfully reserve u, leading to state (R,A) and potentially resulting in state (M,A). In this case, $u$ has been matched by t1, and t0 will stop processing $(u, v)$ because

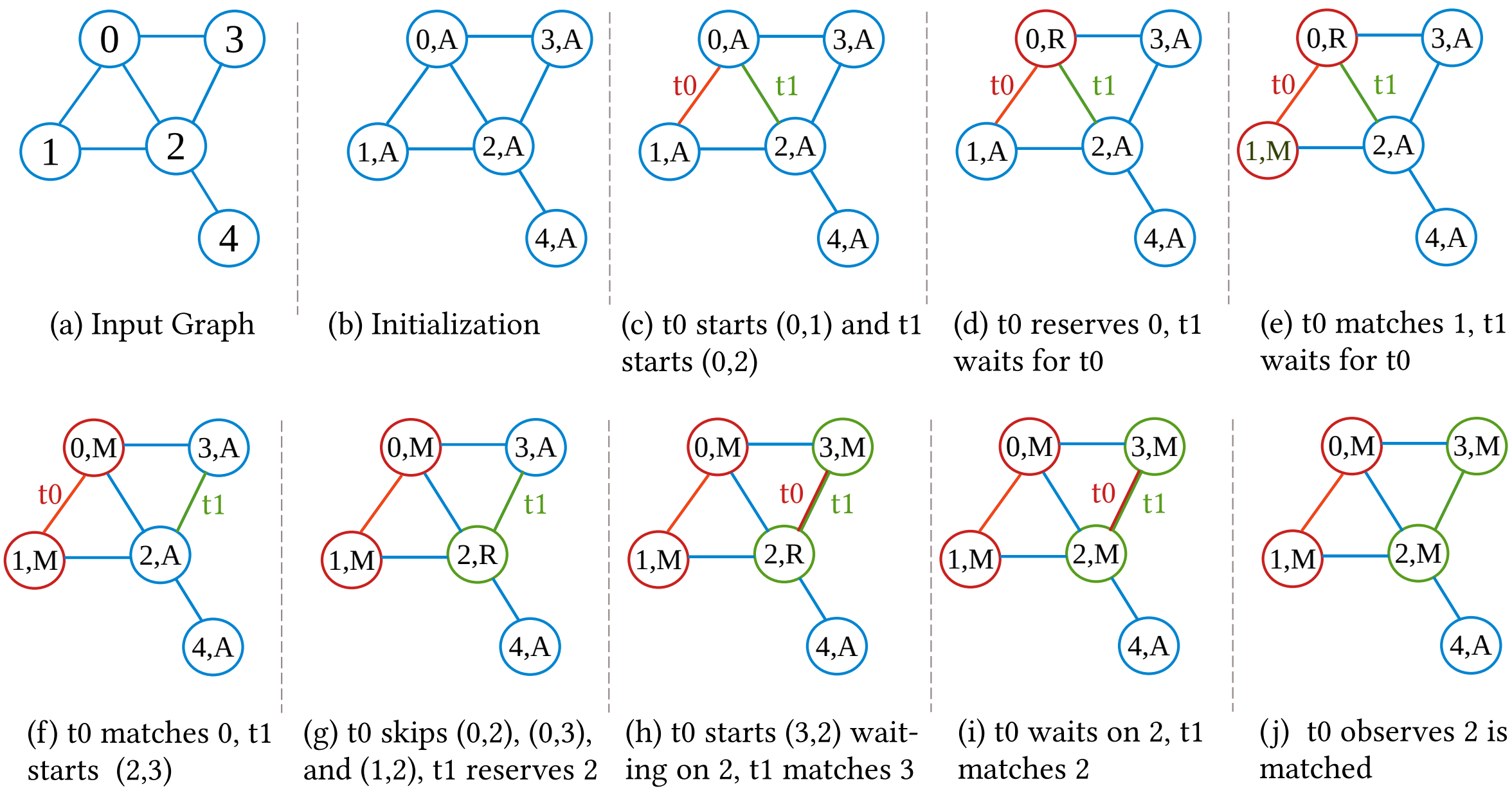

(a) Input Graph (b) Initialization (c) t0 starts (0,1) and t1 starts (0,2) (d) t0 reserves 0, t1 waits for t0 (e) t0 matches 1, t1 waits for t0

(f) t0 matches 0, t1 starts (2,3) (g) t0 skips (0,2), (0,3), and (1,2), t1 reserves 2 (h) t0 starts (3,2) waiting on 2, t1 matches 3 (i) t0 waits on 2, t1 matches 2 (j) t0 observes 2 is matched the condition in Line 10, Algorithm 1, cannot be satisfied. Similarly, thread t2 (pink line) may match $u$ as the second endpoint of the edge it is processing, directly transitioning from (A,A) to (M,A). In these cases edge $(u, v)$ cannot be a match, because $u$ is already an endpoint of a selected edge.

Fig. 6: Parallel execution of Skipper for an example graph by two threads $t0$ and $t1$ - A: ACC, R: RSVD, M: MCHD.

When $u$ and $v$ are in state (A,A), thread t3 (green line) may match $v$ as the second endpoint of the edge it is processing, leading to state (A,M). Similarly, t4 (dashed blue-green line) may successfully match $v$ when t0 has reserved u, i.e., in state (R,A). In this case, t4 leads to state (R,M), and then t0 executes Lines 17–18, releasing $u$ and making it accessible to other threads. In these two cases, $v$ has been matched for another edge, and $(u, v)$ cannot be selected.

**Processing A Whole Graph**. Figure 6 illustrates the parallel execution of Skipper for an example graph. Initially, all vertices are in the ACC state (Figure 6.b). Threads $t0$ and $t1$ start processing from edges $(0, 1)$ and $(0, 2)$, respectively (Figure 6.c). In competition for reserving vertex 0, $t0$ wins (Figure 6.d) and $t1$ waits to observe a non-RSVD state on this vertex. $t0$ matches 1 (Figure 6.e) and changes the state of 0 to $MCHD$ (Figure 6.f). $t1$ observes this and starts processing edge $(2, 3)$ and reserves 2 (Figure 6.g). $t0$ starts processing $(3, 2)$ and by observing 2 has been reserved by another thread, waits on 2 while $t1$ matches 3 (Figure 6.h). Then, $t1$ matches 2 (Figure 6.i) and $t0$ observes this and stops processing $(3, 2)$ (Figure 6.j).

## V. Analysis

### A. Linearizability

We show that any concurrent execution of Skipper is equivalent to a sequential execution by assigning a Linearization Point (LP) to each match operation:

- For a *successful* match of an edge, we choose successful inner CAS at Line 14 as LP for both endpoints of that edge. That CAS is an atomic, irrevocable event that permanently makes $v$ unavailable by transition to state MCHD. Since other threads cannot intervene execution of Line 15, we attribute $LP_u$ at that CAS.
- In matching an edge that ultimately *fails*, the LP is the first action that irrevocably determines failure:
  - A failed CAS at Line 11 (the attempt to reserve $u$) or at Line 14 (the attempt to match $v$), or
  - A plain read at Line 10, that observes $state_v =$ MCHD or $state_u =$ MCHD. If the plain read is transient, the subsequent failing CAS serves as the LP.

Successful CASes on the same vertex are mutually exclusive and the MCHD state is permanent. This ensures that no two successful LPs can claim the same vertex, thereby satisfying the *safety* requirement of the algorithm.

Moreover, every failed attempt observes MCHD or a losing CAS at its LP. Therefore, every unmatched edge has at least a MCHD endpoint, satisfying the *maximality* requirement.

Since each LP is an internal atomic action of its own operation, it becomes observable to other threads instantaneously. Consequently, the total order of LPs *preserves the real-time order* of the original execution.

These facts together establish the linearizability of Skipper.

### B. Complexity

Each edge traversal does $O(1)$ base work, and assuming $t$ threads, CAS retries can be charged for at most the $O(t|V|)$ vertex state transitions. Therefore, the worst case total work is $O(|E| + t|V|)$.

Under a fair scheduler with thread-dispersed locality-preserving scheduling (see below), the probability that any two of $t$ threads experience JIT conflicts on a given vertex is $\lambda = t/|V|$ and $t \ll |V|$. By the Poisson approximation, the probability for two or more threads to make JIT conflicts is $\Theta(\lambda^2)$, which is $\ll 1$ and negligible. Consequently, **JIT conflicts are rare** and **the expected total work** is $O(|E|+|V|)$.

The expected parallel depth is $O(\lambda) = O(t/|V|)$, which simplifies to $O(1)$ when $t \ll |V|$.

Note that the scheduler's thread-dispersed, locality-preserving design (Section IV-C) handles both low- and high-locality graphs:

- For the low- or randomized-locality inputs, the random ordering of vertices makes the probability that two or more threads contend for the same vertex $\Theta(\lambda^2)$, making JIT conflicts rare.
- For high-locality inputs, where consecutive vertex IDs tend to be connected, the scheduler spreads each thread's blocks across the graph, ensuring concurrent threads operate on independent neighborhoods while each thread processes its own blocks sequentially.
  Consequently, the probability of a JIT conflict is bounded by $\Theta(\lambda^2)$, which is $o(1)$, asymptotically smaller than any constant because $\lambda = t/|V| \ll 1$.

In other words, both strong locality and weak locality/randomization are beneficial for reducing JIT conflicts under the thread-dispersed locality-preserving scheduler.

Our empirical results in Section VI-E also show that JIT conflicts occur infrequently on both real-world and synthetic graphs, confirming the practical effectiveness of the scheduler.

### C. Comparison to Previous Work

**Work Complexity & Efficiency**. In Section II-D, we discussed state-of-the-art MM algorithms, which have a work complexity linear in the number of edges, similar to Skipper. However, Skipper requires *exactly one pass over the edges*. Moreover, Skipper avoids costly operations such as graph pruning and randomization. Our evaluation in Section VI-C shows that SIDMM incurs 17–27 memory accesses per edge, whereas Skipper requires only 1.2–3.4. Additionally, SIDMM exhibits 15.4 times as many L3 misses as SGMM, while it is 1.0 times for Skipper.

**Conflict Resolution**. EMS-based MM algorithms accumulate conflicts and resolve them in a separate step (Section II-D). This increases the number of conflicts and the overhead for subsequent iterations. In contrast, Skipper resolves conflicts *instantaneously* using its JIT conflict resolution method, facilitating a single pass over the graph without incurring graph topology operations.

When a JIT conflict happens, Skipper does not require additional memory space or memory writes; it *simply waits a few cycles* to observe the final state of the vertex.

**Synchronization Overhead**. The iterative nature of EMS-based MM algorithms requires synchronization between threads. In contrast, Skipper performs a single pass over the graph edges, allowing it to operate in the Asynchronous Parallel Random Access Machine (**APRAM**) model. Skipper is also **incremental in expectation**, assuming no thread can remain in `RSVD` state in an unbounded steps.

**Simplicity**. One of the Skipper's crucial features is its simplicity, which allows it to be implemented in a few tens of lines in any graph framework. In contrast, previous works require thousands lines of source code, making it difficult to implement in other frameworks.

**No Tuning Parameters**. Unlike previous works, which require empirically tuned parameters (e.g., granularity size in prefix batching or the number of samples in sampling procedure), Skipper achieves its optimal performance without any tuning parameters.

**Non-Deterministic Output**. As explained in Section II-D, EMS algorithms such as SIDMM and PBMM are deterministic, i.e., producing the same output given the same input priority array and sampling seeds. In contrast, Skipper resolves JIT conflicts instantaneously, which makes it non-deterministic. Our evaluation reveals minor variations in the size of the output (i.e., the $matches$ array) across different executions.

**Leveraging Locality and Atomic Memory Accesses**. In contrast to previous works whose optimal performance is guaranteed by randomization, Skipper exploits the graphs' inherent locality to accelerate its execution (Section V-B). Skipper also utilizes atomic Compare-And-Swap memory operations to implement its state transition mechanism, ensuring thread-safe execution without significantly degrading performance[4].

**Input Format & Symmetrization**. Skipper does not imply a restriction on the input format and the input can be provided (i) as a list of edges in coordinate format or (ii) in a compact graph format such as Compressed Sparse Row (CSR), where edges are accessed as neighbors of one of their endpoints. Moreover, in vertex-based graph formats such as CSR, Skipper does not require an edge to be present in the graph topology for both of its endpoints. This omits the symmetrization overhead for directed graphs as a required preprocessing step.

## VI. Evaluation

### A. Experimental Setup

We conduct experiments on a machine equipped with 2 Intel Xeon 6438Y+ CPUs, featuring 64 cores, 64 threads, 2TB DDR5 memory, and running Debian 12.0. The processor has a single-core turbo frequency of 4.0 GHz and all-core turbo frequency of 2.1 GHz.

Table I, Columns 1–4, show the characteristics of the graph datasets. The datasets comprise different graph types, including Web graphs [11–17], Bio graphs [18], Social networks [19], and Synthetic graphs [20]. The graphs have been processed using their published vertex ordering. As discussed in Section V-B, Skipper's performance is independent from vertex ordering.

We implemented Skipper in LaganLighter [21] using `OpenMP`, `PAPI` [22], `libnuma`, and ParaGrapher with PG-Fuse[5] [23]. Skipper's source code is available online[6]. We evaluate Skipper against the implementation of SIDMM algorithm in GBBS [7][7] explained in Section II-D. We used `gcc` 14.0.1 compiler with the `-O3` optimization flag.

[4] "Atomics are not that bad anymore" [10].
[5] https://github.com/MohsenKoohi/ParaGrapher
[6] https://github.com/MohsenKoohi/LaganLighter

TABLE I: Performance of Skipper in comparison to SIDMM in GBBS[7] (Section II-D). Execution time has been measured after loading the entire topology data of the symmetrized graph into memory. SIDMM could not complete processing `wdc12` in 240 seconds.

| Dataset | | | | Performance (sec) | | Speedup |
|---|---|---|---|---|---|---|
| Name | Type | \|V\| | \|E\| | SIDMM | Skipper | |
| twitter10 | Social | 41.7M | 2.4G | 0.60 | 0.12 | 5.1 |
| g500 | Synth. | 536.9M | 17.0G | 6.6 | 0.68 | 9.6 |
| msa10 | Bio | 1.8G | 41.7G | 24.9 | 5.1 | 4.9 |
| clueweb12 | Web | 978.4M | 74.7G | 23.2 | 2.7 | 8.7 |
| wdc14 | Web | 1.7G | 123.9G | 41.1 | 2.6 | 15.6 |
| eu15 | Web | 1.1G | 161.1G | 26.6 | 3.3 | 8.0 |
| wdc12 | Web | 3.6G | 224.5G | – | 9.4 | – |

### B. Performance

Table I, Columns 5–6, report the execution time of SIDMM and Skipper algorithms in seconds. Column 7 shows the speedup of Skipper compared to SIDMM. Execution times are in seconds, measured after the entire topology data of the symmetrized graph loaded into memory, i.e., the execution times do not include symmetrization and loading time. SIDMM could not finish processing `wdc12` in 240 seconds.

Table I shows that **Skipper achieves a speedup of 4.9–15.6 times, with a geometric mean of 8.0 times**. Skipper performs a single pass over graph and does not require graph pruning or randomization, which facilitates significant speedup for it. For graphs larger than 50 billion edges, Skipper achieves a geometric mean speedup of more than an order of magnitude.

### C. Work Efficiency

To assess the work efficiency of MM algorithms as memory-intensive algorithms, we evaluate two key metrics: (i) the number of memory accesses (load and store instructions) and (ii) the number of Level 3 (L3) cache misses for SIDMM and Skipper as parallel algorithms compared to SGMM as the sequential reference algorithm.

Figure 7 compares the number of memory accesses for SGMM (sequential, single threaded) and SIDMM and Skipper (parallel, 64-threaded). Values are normalized to the number of graph edges. Note that memory accesses include all accesses, regardless of cache hits or misses. Figure 7 shows that SGMM performs 0.3–0.8 times memory accesses per edge, which increases to 16.7–26.9 for SIDMM. In contrast, Skipper requires only 1.2–3.4 memory accesses per edge. **The geometric mean of memory accesses per edge for SIDMM is 21.0, while for Skipper it is 2.1**.

[7]https://github.com/ParAlg/gbbs/tree/bd48872/benchmarks/MaximalMatching/RandomGreedy/

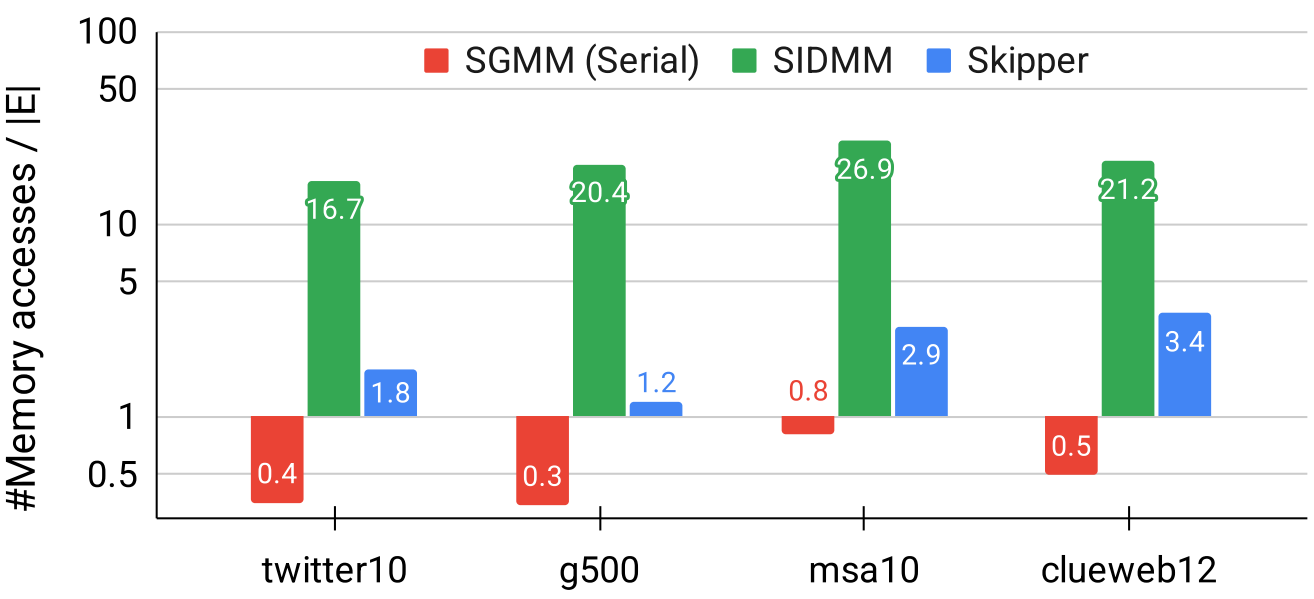

Fig. 7: Number of memory accesses normalized to $|E|$ for SGMM (sequential MM algorithm running on a single thread) and SIDMM and Skipper (running on 64 threads).

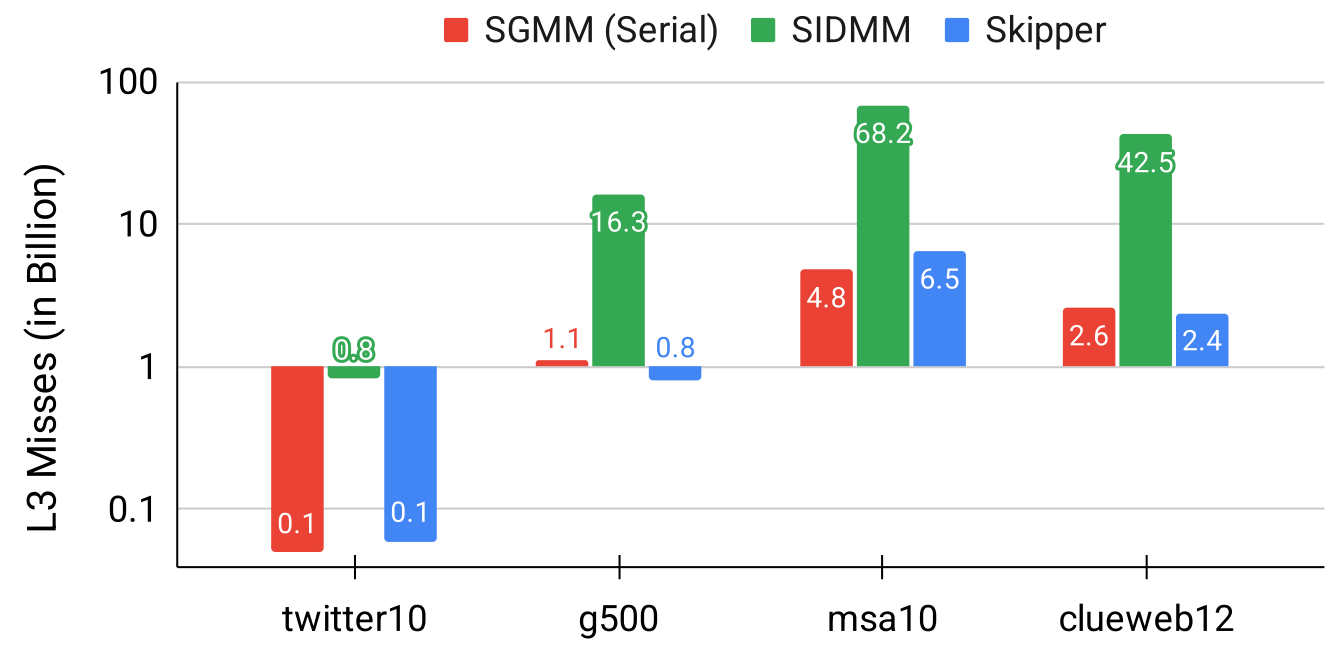

Fig. 8: Number of L3 misses of SGMM (sequential MM algorithm running on a single thread) and SIDMM and Skipper (running on 64 threads).

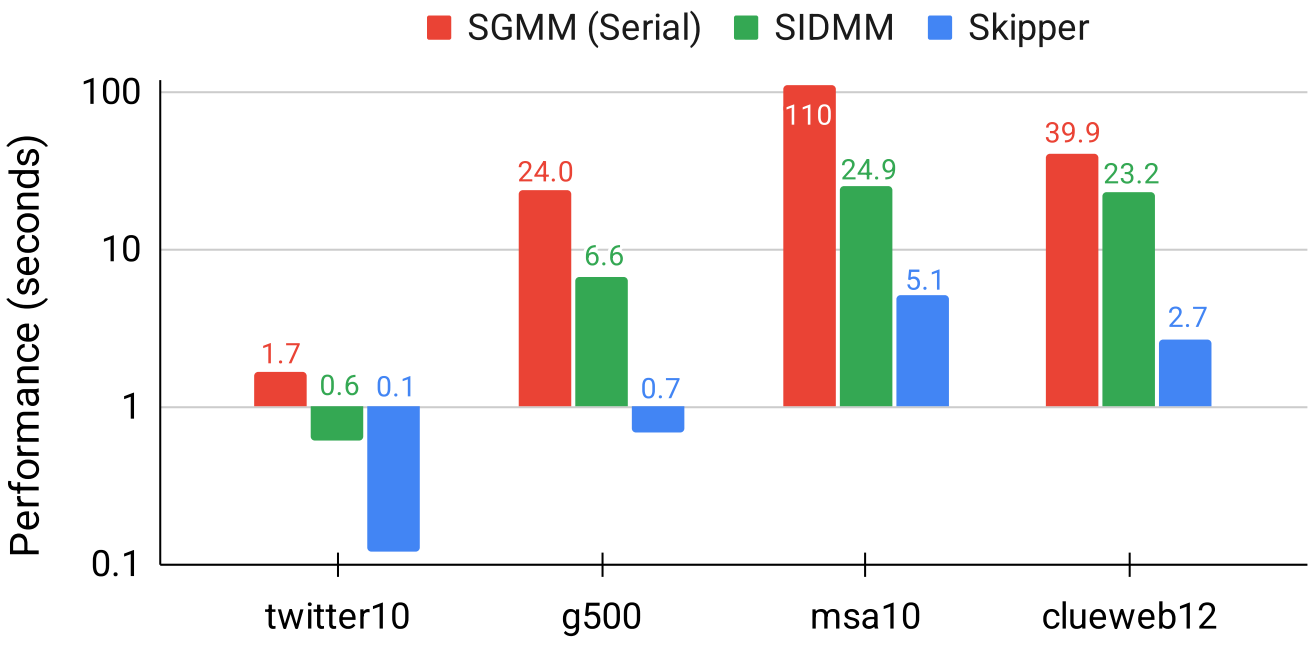

Fig. 9: Performance of SGMM (sequential MM algorithm running on a single thread) and SIDMM and Skipper (running on 64 threads) in seconds.

Figure 8 compares the number of L3 cache misses. It shows that SIDMM incurs 14.2–16.5 times the L3 misses of SGMM, while Skipper incurs 0.7–1.4 times that of SGMM. **The geometric mean of L3 misses relative to SGMM is 15.4 for SIDMM and 1.0 for Skipper**. Notably, for `g500` and `clueweb12`, Skipper's L3 misses are fewer than SGMM's.

Figure 9 compares the execution times of SGMM, SIDMM, and Skipper. SGMM runs on a single thread, while SIDMM and Skipper use 64 threads. Figure 9 shows that SIDMM's 64-threaded execution is 1.7–4.5 times faster than SGMM's single-threaded execution (geometric mean of 3.0 times), while Skipper's 64-threaded execution is 14.0–35.2 times faster (geometric mean of 20.0 times). This highlights the impact of reduced excess work (including multiple passes over the graph, thread synchronization and graph pruning) in Skipper compared to SIDMM.

Furthermore, the similar number of L3 misses between Skipper and SGMM, combined with Skipper's 20.0 times speedup, indicates that **Skipper achieves a 20.0 times speedup over SGMM by maintaining work efficiency close to the sequential algorithm.**

### *D. Parallelization Gain & Serial Slowdown*

In parallel algorithms, the **Strong Scalability Test** assesses performance as the number of threads increases, expecting a linear reduction in execution time proportional to the number of threads. However this evaluation relies on:

*Assumption 1*: "More cores" equals to "proportional increase in all computational resources".

For computation-intensive parallel algorithms, the minimum requirements of this assumption are: (i) L1 cache size is sufficient to include the working dataset for each core, (ii) threads do not access each other's data, and (iii) core frequency remains constant with increasing active cores.

However, for memory-intensive parallel algorithms on NUMA architectures, the assumption may not remain valid as:

- Increasing the number of active NUMA nodes changes the memory access costs, potentially reducing memory efficiency depending on the architecture, algorithm, and input dataset.
- In single-socket experiments, constant memory channels and L3 cache mean that not all resources scale with the increasing number of cores.

Therefore, in memory-intensive algorithms, treating "more cores" as equivalent to "more computational resources" is not a valid assumption and it is normal to observe non-linear changes (either worsening, or less commonly improving) in runtime as core count increases.

*Assumption 2*: Single-core execution of a parallel algorithm is the most efficient sequential execution, making it a fair base for measuring efficiency of parallel algorithms. This assumption is often incorrect for memory-intensive parallel algorithms, as the best sequential algorithm is usually much faster than single-threaded execution of a parallel algorithm.

Due to these limitations, the scalability tests are not well-suited for memory intensive parallel algorithms. Instead, we use the following metrics to measure the parallelization capability:

- **Parallelization Gain** (or Parallel Speedup): comparing the performance of a parallel algorithm to the best sequential algorithm. Parallelization Gain is defined as $t_s/t_p$, where $t_s$ and $t_p$ are the execution time of the sequential and parallel algorithms, respectively.
  Note that Parallelization Gain has limitations. Similar to the above reasoning in Assumption 1, we may not observe a parallel algorithm running on $t$ threads achieving $t$-fold speedup compared to the sequential algorithm.

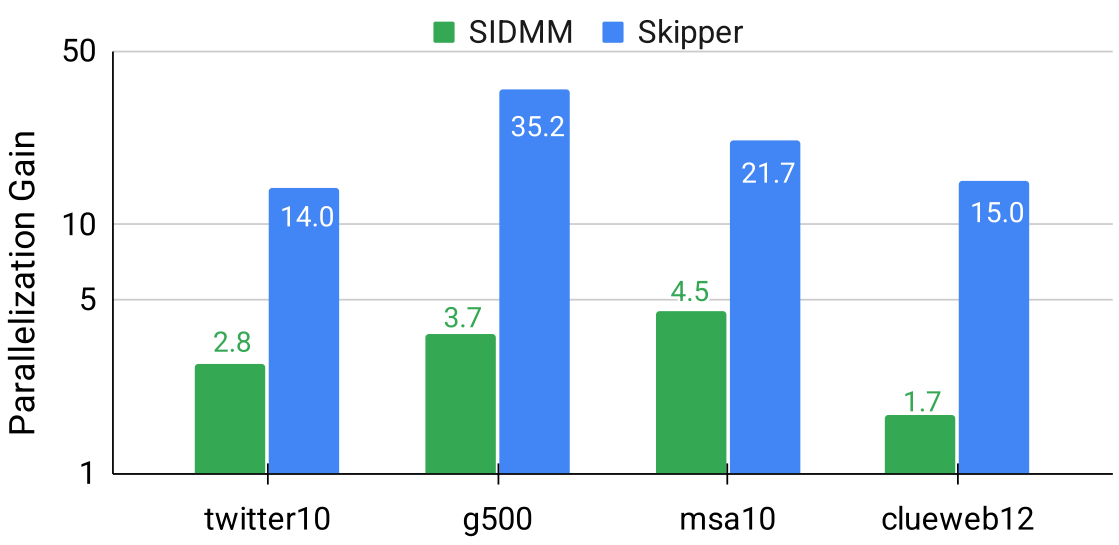

Fig. 10: Parallelization gain of SIDMM and Skipper (running on 64 threads) relative to SGMM (sequential). Parallelization gain is defined as $t_s/t_p$, where $t_s$ and $t_p$ are the execution time of the sequential and parallel algorithms, respectively.

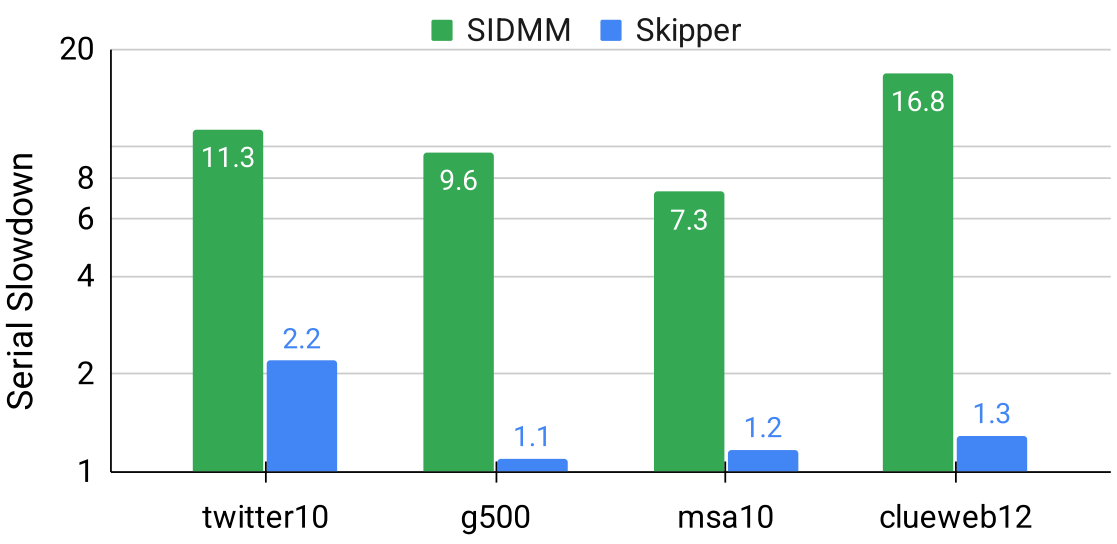

Fig. 11: Serial Slowdown of SIDMM and Skipper relative to SGMM. All algorithms are running on a single thread. Serial slowdown is defined as $t_{p,1t}/t_s$, where $t_{p,1t}$ is the execution time of the parallel algorithm running on a single thread and $t_s$ is the execution time of the sequential algorithm.

- **Serial Slowdown**: comparing single-threaded execution of a parallel algorithm over the best sequential algorithm. It is defined as $t_{p,1t}/t_s$, where $t_{p,1t}$ is the execution time of the parallel algorithm running on a single thread and $t_s$ is the execution time of the sequential algorithm.
  Serial Slowdown reflects how much slowdown is applied by a single-threaded execution of a parallel algorithm. A lower Serial Slowdown usually indicates that the parallel algorithm has less overhead and is more efficient. While Serial Slowdown can be a good indicator of parallelization capability, it may not be universally applicable.

Figure 10 shows the Parallelization Gain of SIDMM and Skipper relative SGMM. The results show that with using 64 threads, SIDMM achieves a parallelization gain of 1.7–4.5, whereas Skipper achieves a significantly higher parallelization gain of 14.0–35.2 .

Figure 11 shows the Serial Slowdown of SIDMM and Skipper relative to SGMM. The results indicate that the single-threaded execution of SIDMM incurs a slowdown of 7.3–16.8 times (geometric mean: 10.7). In contrast, Skipper's single-threaded execution incurs a significantly lower slowdown of 1.1–2.2 times (geometric mean: 1.4 times).

### *E. JIT Conflicts*

Table II presents the characteristics of Skipper's JIT conflicts when running on 64 and 16 threads. A conflict is formally defined as a failing $CAS$ in Algorithm 1 (Lines 11 and 14).For each edge, the number of conflicts is the sum of conflicts for both edges. The table provides the following key information:

- Column 3 shows the maximum number of conflicts over all edges of a graph.
- Column 4 shows the total number of conflicts occurred while processing the graph.
- Column 5 indicates the number of edges that have experienced at least one conflict.
- Column 6 shows the average number of conflicts for each conflicting edge.
- Columns 7–16 show the distribution of edges by the number of conflicts. For example, Row 3, Columns 7–16 indicate that for `twitter10`, 55 edges have experienced a single conflict, one edge has experienced 17–32 conflicts, and two edges have experienced 33–64 conflicts.

TABLE II: Analyzing JIT conflicts in Skipper running on 64 and 16 threads. A conflict happens when a $CAS$ operation in Algorithm 1 fails (Lines 11 and 14). **Max cnf per edge**: The maximum number of conflicts encountered by an edge. **Total cnfl**: The total number of conflicts across all edges. **#Edges exp. cnf**: The number of edges that experienced conflicts. **Avg. cnf per edge**: The average number of conflicts per those edges that experienced conflicts. **Conflict distribution**: The distribution of edges categorized by the number of conflicts.

| Dataset | Num threads | Max cnf per edge | Total cnfl | #Edges exp. cnf | Avg. cnf per edge | Conflict distribution | | | | | | | | | |
|---|---|---|---|---|---|---|---|---|---|---|---|---|---|---|---|
| | | | | | | 1 | 2 | 3–4 | 5–8 | 9–16 | 17–32 | 33–64 | 65–128 | 129–256 | >256 |
| **twitter10** | 64 | 53 | 179 | 58 | 3.1 | 55 | | | | | 1 | 2 | | | |
| | 16 | 23 | 37 | 14 | 2.6 | 12 | 1 | | | | 1 | | | | |
| **g500** | 64 | 54 | 148 | 40 | 3.7 | 33 | 2 | | 1 | 2 | 1 | 1 | | | |
| | 16 | 38 | 44 | 7 | 6.3 | 6 | | | | | | 1 | | | |
| **msa10** | 64 | 410 | 44371 | 26360 | 1.7 | 15485 | 8771 | 1281 | 567 | 233 | 11 | 9 | 2 | | 1 |
| | 16 | 27 | 89 | 19 | 4.7 | 10 | 4 | 2 | | 1 | 2 | | | | |
| **clueweb12** | 64 | 6 | 28 | 18 | 1.6 | 12 | 5 | | 1 | | | | | | |
| | 16 | 1 | 1 | 1 | 1.0 | 1 | | | | | | | | | |
| **wdc14** | 64 | 36 | 5902 | 877 | 6.7 | 385 | | 1 | 175 | 280 | 34 | 2 | | | |
| | 16 | 29 | 166 | 22 | 7.5 | 10 | | 1 | 3 | 6 | 2 | | | | |
| **eu15** | 64 | 22 | 29 | 8 | 3.6 | 7 | | | | | 1 | | | | |
| | 16 | 1 | 5 | 5 | 1.0 | 5 | | | | | | | | | |
| **wdc12** | 64 | 26 | 58 | 32 | 1.8 | 30 | 1 | | | | 1 | | | | |
| | 16 | 29 | 55 | 6 | 9.2 | 4 | | | | | 2 | | | | |

To collect the data in Table II, we executed Skipper for five times over each graph and selected the execution with the largest number of edges that have experienced conflicts.

Table II shows that the maximum number of conflicts for a conflicting edge has been 410. Also, the maximum average number of conflicts for each conflicting edge has been 9.2 and the maximum total conflicts is fewer than 45k.

Among all datasets, `msa10` exhibits the highest JIT conflict ratio, (conflicting edges$/|E|$). Specifically, the number of conflicting edges for this graph is $< 27k$ out of 42 billion edges. This gives a conflict ratio less than 0.1%, empirically confirming **JIT conflicts are rare**.

The conflicts distribution shows that the majority of conflicting edges have experienced less than 16 conflicts. Additionally, Table II shows that reducing the number of threads reduces the number of conflicts.

## VII. Conclusion

This paper introduces Skipper, an asynchronous Maximal Matching algorithm that processes the entire graph in a single pass over the edges. Unlike the previous works that accumulate the conflicts and process the graph iteratively, Skipper introduces a Just-In-Time (JIT) conflict resolution method. Skipper operates in the APRAM model and its computation progresses without requiring the entire graph to be accessed at once.

Our evaluation, conducted on graphs with up to 224 billion edges, demonstrates that Skipper achieves a significant speedup of 8.0 times (geometric mean). Notably, for graphs larger than 50 billion edges, Skipper achieves a speedup of over an order of magnitude (geometric mean).

## Source Code

The source code of Skipper is available on https://github.com/MohsenKoohi/LaganLighter.

## Acknowledgements

The computational resources of this work has been partially provided by NI-HPC (UKRI EPSRC grant EP/T022175/1).

We used a non-training version of Llama 4 to enhance grammar and editing in this paper.